\documentclass[sn-standardnature]{sn-jnl}
\usepackage{lineno}

\jyear{2021}%

\theoremstyle{thmstyleone}%
\theoremstyle{thmstyletwo}%

\theoremstyle{thmstylethree}%

\begin{document}

\title[Magnetic skyrmion lattices in a novel two-dimensional twisted bilayer magnet]{Magnetic skyrmion lattices in a novel two-dimensional twisted bilayer magnet}


\author*[1,2]{\fnm{Fawei} \sur{Zheng}}\email{fwzheng@bit.edu.cn}

\affil[1]{\orgdiv{Centre for Quantum Physics, Key Laboratory of Advanced Optoelectronic Quantum Architecture and Measurement (MOE)}, \orgname{School of Physics, Beijing Institute of Technology}, \orgaddress{\city{Beijing}, \postcode{100081}, \country{China}}}
\affil[2]{\orgdiv{Beijing Key Lab of Nanophotonics $\&$ Ultrafine Optoelectronic Systems}, \orgname{School of Physics, Beijing Institute of Technology}, \orgaddress{\city{Beijing}, \postcode{100081}, \country{China}}}


\abstract{Magnetic skyrmions are topologically protected spin swirling vertices, which are promising in device applications due to their particle-like nature and excellent controlability. Magnetic skyrmions have been extensively studied in a variety of materials and were proposed to exist in the extreme two-dimensional limit, i.e., in twisted bilayer CrI$_3$ (TBCI). Unfortunately, the magnetic states of TBCIs with small twist angles are disorderly distributed ferromagnetic (FM) and antiferromagnetic (AFM) domains in recent experiments, and thus the method to get rid of disorders in TBCIs is highly desirable. Here we use intralayer exchange interactions up to the third nearest neighbors without empirical parameters and very accurate interlayer exchange interactions to study the magnetic states of TBCIs. We propose the functions of interlayer exchange interactions obtained using first-principles calculations and stored in symmetry-adapted artificial neural networks. Based on them, the subsequent Landau-Lifshitz-Gillbert equation calculations explain the  disorderly distributed FM-AFM domains in TBCIs with small twist angles and predict the orderly distributed skyrmions in TBCIs with large twist angles. This novel twisted two-dimensional bilayer magnet can be used to design memory devices, monochromatic spin wave generators and many kinds of skyrmion lattices.}

\maketitle

\section{Introduction}

Since the successful construction of twisted bilayer graphene\cite{TBG,TBG2}, rich properties of twisted bilayer systems have been demonstrated, such as correlation-driven insulators\cite{kerelsky2019maximized,po2018origin}, superconductivity\cite{yankowitz2019tuning,wu2018theory},  ferromagnetism\cite{sharpe2019emergent,lu2019superconductors,serlin2020intrinsic}, moiré excitons\cite{tran2019evidence,jin2019observation,seyler2019signatures}, and topological states\cite{nuckolls2020strongly,wu2021chern}. Recently, there has emerged the twisted bilayer CrI$_3$ (TBCI) with intrinsic magnetism.  The magnetic state of a CrI$_3$ bilayer can be tuned by electric field\cite{huang2018electrical,jiang2018electric,morell2019control,xu2020electric}, external pressure\cite{song2019switching,Shan,yang2019van,subhan2019pressure}, and charge doping\cite{jiang2018controlling,PhysRevB.101.041402,ghosh2021overcoming}. Theoretical studies have predicted that it also depends on the stacking structure\cite{Sivadas,Soriano,Ji,Leon2022ATS},  and a twisting may bring periodic magnetization domains with complex spin texture\cite{wang2020stacking,TBCI1,soriano2022domain}; additional Dzyaloshinskii-Moriya (DM) interactions may further stabilize various magnetic skyrmions\cite{TBCI2,TBCI3,TBCI4,ghader2021whirling}. Rich phase diagrams with noncollinear spin configurations were obtained\cite{Hejazi2020PNAS,Ray2021PRB}. Compared with magnetic skyrmions in alloys\cite{MagneticSkyrmion2,MagneticSkyrmion3,MagneticSkyrmion4,Fert2017magnetic}, the magnetic skyrmions in TBCIs are much thinner, which reach the two-dimensional limit and open up the field of spintwistronics.

\begin{figure}
\centering
\includegraphics[width=\linewidth]{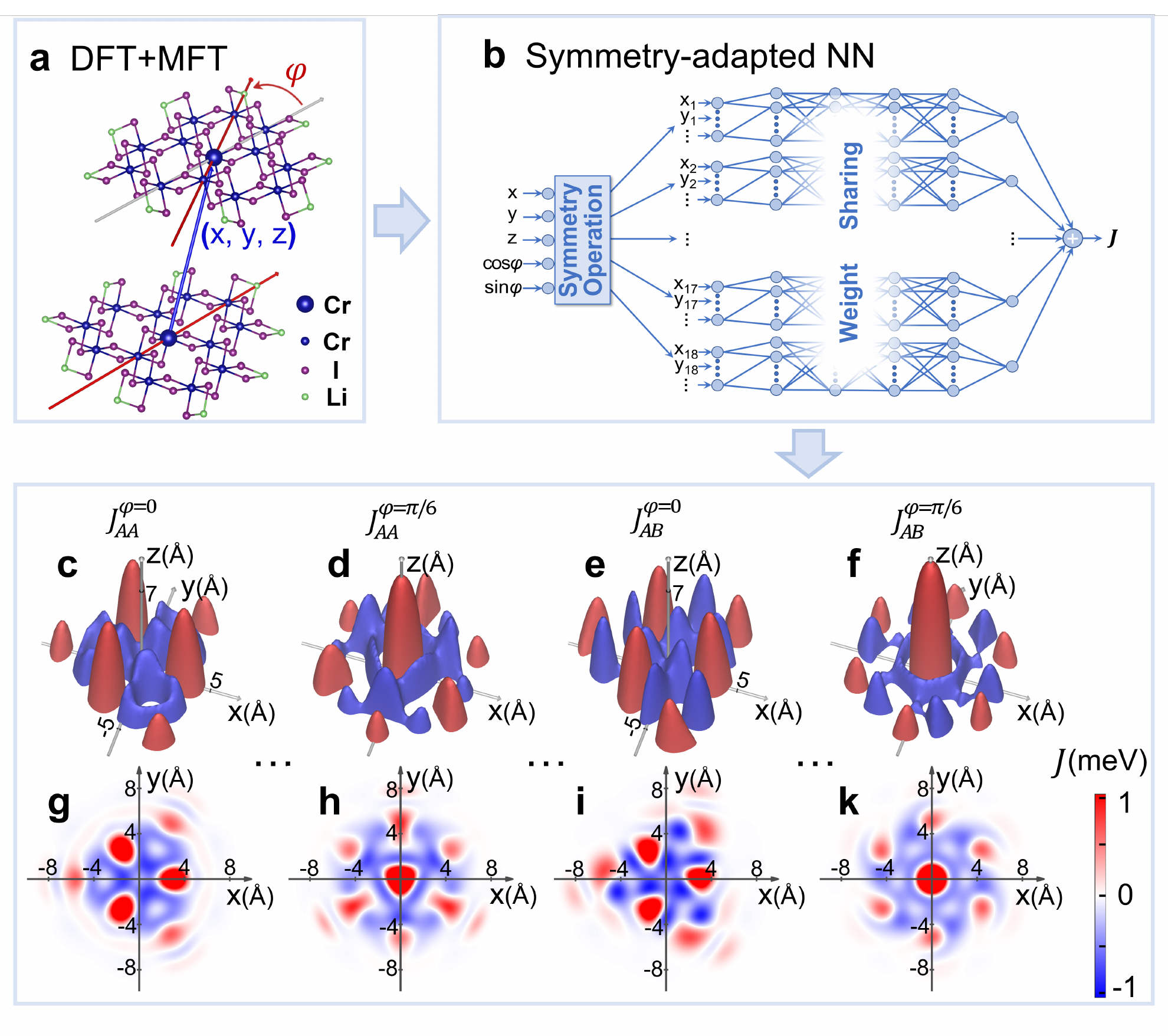}
\caption{(a)The twisted bilayer CrI$_3$ cluster model with the central Cr atoms shown in large balls to highlight their positions; (b) the structure of a SANN; (c,d) the isosurfaces of $J_{AA}$ for $\varphi=$0 and $\pi/6$, respectively; (e,f) that of $J_{AB}$. The isosurfaces for the red and blue colors are 0.41 meV and -0.41 meV, respectively, where the positive (negative) value signs the FM (AFM) exchange interaction. The beginning of $z$ axis at the bottom is 6.5\AA. The corresponding coss section views of $J_{AA}$ and $J_{AB}$ at $z=$6.62 \AA\, are shown in panel (g-k).}
\label{fig1}
\end{figure}

In previous experiments, TBCIs with small twist angles show the coexistence of interlayer ferromagnetic (FM) and antiferromagnetic (AFM) states, but their magnetic domains are disorderly distributed\cite{TBCIexp1,TBCIexp2}. Therefore, experimentalists turned to the study of twisted multilayer CrI$_3$, such as twisted double bilayer CrI$_3$\cite{TBCIexp3,Xie2022evidence,cheng2022electrically} and twisted double trilayer CrI$_3$\cite{TBCIexp2}, since stacking more layers of CrI$_3$ suppresses the disorder. Then, orderly distributed magnetic domains were detected in twisted double trilayer CrI$_3$\cite{TBCIexp3}. However, the method for getting orderly distributed magnetic skyrmions in the much simpler TBCI is still unknown. In this work, we find a solution by extensively studying the magnetic states of TBCIs with arbitrary twist angles. Two kinds of TBCI  were  found  having the coexistence of FM-AFM states. The first kind of TBCI has a small twist angle, and it has been well studied in literature; the other kind is new, whose twist angle is around $60^\circ$. These two kinds of TBCI have quite different interlayer exchange fields, which explains the disorderly distributed FM-AFM domains in experiments and suggests the existence of orderly distributed skyrmions in the new kind of TBCI. 

\section{Results and Discussion}
\subsection{The Magnetic Exchange Interaction in TBCI}
The magnetic exchange interactions in a TBCI are composed of intralyer and interlayer interactions. The interlayer interaction determines the interlayer magnetic order, and thus is the key to simulating magnetic states of TBCIs. In literature,  the interaction is usually approximated by using the interlayer interactions in the non-twisted bilayer CrI$_3$, which  are only applicable to the study of TBCIs with small twist angles. In this work, we used the accurate interlayer magnetic exchange interaction instead of the non-twist approximation. The interlayer magnetic exchange interaction beween two Cr atoms depends on their in-plane relative position ($x=rcos\theta$, $y=rsin\theta$), interlayer distance ($z$), and the twist angle ($\varphi$). Since the Cr atoms in a CrI$_3$ layer form a honeycomb lattice consisting of two sublattices (named as A and B), the  interlayer magnetic exchange interaction $J(r,\theta,z,\varphi)$ has four different kinds, that are $J_{AA}$, $J_{AB}$, $J_{BA}$, and $J_{BB}$. The first and second footnotes denote the sublattices of the two Cr atoms. For instance, $J_{AB}$ describes the magnetic interaction between a Cr atom in the sublattice-A of lower layer and another Cr atom in the sublattice-B of upper layer. These four $J$-functions are not independent, since $J_{BA}$ and $J_{BB}$ can be obtained from $J_{AB}$ and $J_{AA}$ by symmetry constrains: $J_{BB}(r,\theta,z,\varphi)=J_{AA}(r,-\theta,z,-\varphi)$ and $J_{BA}(r,\theta,z,\varphi)=J_{AB}(r,-\theta,z,-\varphi)$. Besides, the symmetry of TBCI provides the following additional constrains (see the Supporting Information for further details on symmetry constrains).

\begin{equation}
\begin{aligned}
J_{AA(B)}(r,\theta,z,\varphi)&=J_{AA(B)}(r,\theta\pm\frac{2\pi}{3},z,\varphi)\\
J_{AA(B)}(r,\theta,z,\varphi)&=J_{AA(B)}(r,\theta,z,\varphi\pm\frac{2\pi}{3})\\
J_{AA}(r,\theta,z,\varphi)&=J_{AA}(r,-\theta+\varphi,z,\varphi) \\
J_{AB}(r,\theta,z,\varphi)&=J_{AB}(r, \theta-\varphi,z,-\varphi) \\
\end{aligned}
\end{equation}

These constrains reduce the calculation of $J$-functions to a smaller variable space.  The unit cells for $\varphi=0$ and $\pi/3$ are very small, thus the values of $J_{AA}$ and $J_{AB}$ can be obtained directly by performing density functional theory (DFT) calculations. In contrast, for most of the remaining twist angles, the  TBCIs have large Mori\'{e} patterns with huge amounts of atoms, which {  prevent} the direct use of DFT calculations.    

To solve the large-cell problem, we designed a cluster model, as shown in Figure \ref{fig1}(a), to calculate the interlayer magnetic exchange interactions for arbitrary twist angles and translations. The cluster model contains two parallel CrI$_3$ plates, each of them has 13 Cr, 48 I, and 9 Li atoms. The radius of the plate is as large as 11.9 \AA. The Li  atoms are used to balance the electron-losing at the plate edge. We have checked the reliability of the cluster models by testing different cluster diameters, choosing different edge structures, and comparing the interlayer exchange interactions with those of periodic models (see the Supporting Information for further details on model testing). The cluster models were set with large amount of different stacking structures, with a 10$\times$10 translation mesh from 0 \AA\, to 9 \AA, eight twist angles in the range of 0$\sim2\pi/3$, and seven interlayer distances (between the two Cr-plane) from 6.55 \AA\, to 7.53 \AA. The total number of stacking structures is 11200 for $J_{AA}$ and $J_{AB}$.

After DFT convergence,  the values of $J_{AA}$ and $J_{AB}$ were obtained by using a magnetic force theory (MFT) method\cite{MFT1,MFT2}. We then further constructed and trained symmetry-adapted artificial neural networks (SANNs), as shown in Figure \ref{fig1}b, to predict $J$-functions for periodic TBCIs with any stacking structures. We have used two SANNs to predict $J_{AA}$ and $J_{AB}$ seperately, since the two functions have different symmetries. The input of a SANN is $(x,y,z,cos\varphi,sin\varphi)$. The twist angle was not used directly; instead we have used $cos\varphi$ and $sin\varphi$ to naturally encode the periodic condition in neural networks. The input data was proliferated into 18 equivalent copies according to the symmetries of the $J$ function. Then the 18 copies were seperately fed into 18 feed-forward neural networks (FFNNs). These FFNNs share the same structure and parameters. Each branch of FFNN has four hidden layers containing 40, 100, 100, and 40 neurons, respectively. At last, the predicted results from the 18 branches were added together to produce the final prediction. The predicted $J_{AA}$ and $J_{AB}$ for two typical twist angles are shown in Figure \ref{fig1}c, where the $J$ values decrease dramatically with an increase of the interlayer distance, and may switch signs by changing twist angles and in-plane relative positions. These $J$-functions from SANNs have been tested by reproducing the interlayer exchange energies of typical CrI$_3$ bilayers\cite{Sivadas,Yu} (see the Supporting Information for further details on interlayer exchange energies). Based on the trained SANNs, the interlayer interaction parameters for a large TBCI can be obtained in very limited calculation consumings.

Due to the strong spin-orbit coupling and the unique atomic structure, the intralayer magnetic exchange interactions of CrI$_3$ are also complex. We have used the four-state method\cite{xiang2013magnetic,PhysRevB.84.224429,ChangsongXu} to obtain the full 3$\times$3 exchange matrices $\mathcal{J}$ up to the third nearest neighbors, as shown in Figure \ref{fig2}a. The exchange interactions for the fourth nearest neighbors were neglected, since the corresponding values of $\mathcal{J}$ were estimated to be smaller than 0.05 meV in MFT calculations. Our calculation results (see the Supporting Information for further details) show that the exchange matrices for the first nearest neighbors and the third nearest neighbors are symmetric, which agree with the inversion symmetries between the two Cr atoms. The exchange matrices for the second nearest neighbors, however, are non-symmteric due to the missing of inversion symmetry. Therefore, $\mathcal{J}_2$ contains DM interaction, wihch may stabilize magnetic skyrmions.

\subsection{The Magnetic Domains in TBCI}

\begin{figure}
\centering
\includegraphics[width=.8\linewidth]{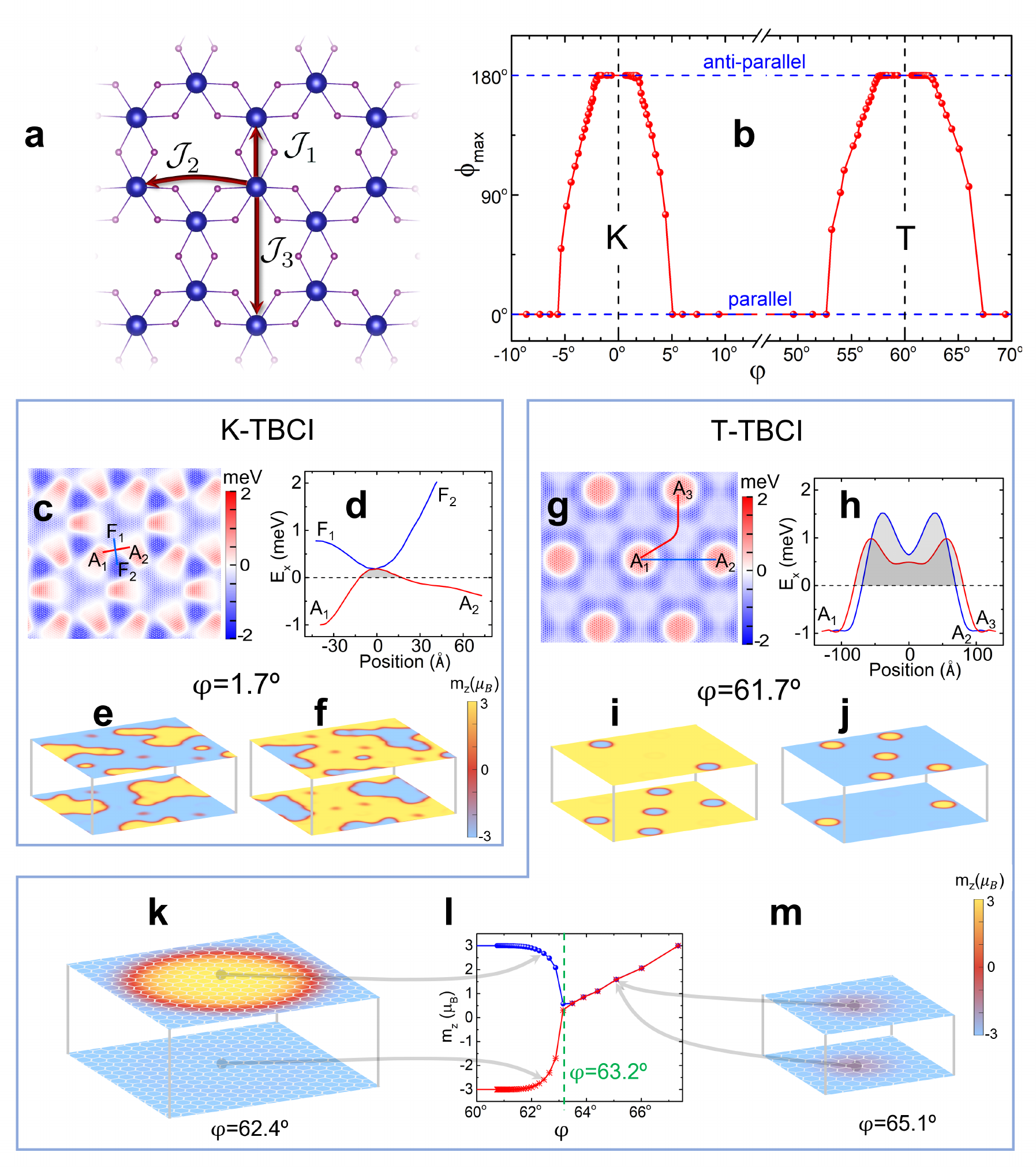}
\caption{(a) The first, second and third nearest neighboring interactions in a CrI$_3$ single layer; (b) the maximum angle difference ($\phi_{max}$) between a magnetic moment in the upper layer and a nearby magnetic moment in the lower layer;  (c) the interlayer exchange field of a K-TBCI with twist angle $\varphi=1.7^\circ$; (d) the interlayer exchange field along  $A_1-A_2$ and $F_1-F_2$ in panel (c); (e,f) two typical magnetic states of the K-TBCI; (g) the interlayer exchange field of a T-TBCI with $\varphi=61.7^\circ$; (f) the interlayer exchange fields along  $A_1-A_2$ and $A_1-A_3$ in panel (g); (i,j) two typical magnetic states of the T-TBCI; (k,m) two different magnetic domains in T-TBCI; (l) the  $m_z$ component at the domain center in the upper and lower layers as a function of twist angle.}\label{fig2}
\end{figure}

After obtaining the interlayer parameters from the SANNs and intralayer parameters from DFT calculations,  we then constructed the full Hamiltonians of TBCIs. The Hamiltonian of a TBCI can be written as:

\begin{equation}
\begin{aligned}
H=\sum_{i>j>0}\mathbf{S}_i^T\mathcal{J}_{ij}\mathbf{S}_j + \sum_{i<j<0}\mathbf{S}_i^T\mathcal{J}_{ij}\mathbf{S}_j + \sum_{i>0,j<0}J_{ij}\mathbf{S}_i\cdot\mathbf{S}_j
\end{aligned}
\end{equation}  

where $\mathbf{S}_i$, whose length is $3/2$, is the spin of the Cr atom at site-$i$. The single ion anisotropy is negelected, since it is much smaller than the anisotropic exchange in CrI$_3$\cite{Lado2017}. We have signed all the Cr atoms in the upper layer by positive indices and those in the lower layer by negative indices. The first and second terms at the right side describes the intralayer exchange interactions for upper and lower layers, respectively.  The matrices $\mathcal{J}_{ij}$ for the first, second and third nearest neighbors were obtained by DFT calculations, and the other intralayer magnetic exchange interactions were neglected. The matrices $\mathcal{J}_{ij}$ are different in different coordinate systems. A TBCI with nonzero {  twist angle} $\varphi$ contains a rotated CrI$_3$ layer, where the matrix $\mathcal{J}_{ij}$ were transformed to the new coordinate system by the rotation matrix, $\mathcal{J}'_{ij}=\mathcal{R}^{-1}(\varphi)\mathcal{J}_{ij}\mathcal{R}(\varphi)$. The third term describes the interlayer exchange interactions, whose value was predicted by the SANNs. It is well known that the local interlayer distance varies depending on the position in the Mori\'{e} pattern and changes the strength of interlayer exchange interactions. We adopted a scheme to approximately estimate the interlayer distances by using a dipole-force based interpolation, and predicted more realistic interlayer exchange interactions. More details can be found in the Supporting Information.

We then calculated ground magnetic states by performing the Landau-Lifshitz-Gillbert (LLG) equation {  calculations}. The calculation results show that the ground states are FM in most cases, except for the TBCIs with $\varphi$ around $0^\circ$ or $60^\circ$, which show the co-existence of FM and AFM states. In order to identify the FM and AFM domains, we have investigated the angle between the magnetic moment of a Cr atom in the upper layer and that of neighboring Cr atoms in the lower layer. The minimum angle is always $0^\circ$, which shows that the FM domains always exist. The maximum angle as a function of the twist angle is shown in Figure \ref{fig2}b, where the angle increases significantly when $\vert\varphi\vert<5.1^\circ$ or $\vert\varphi-60^\circ\vert<7.3^\circ$, and reaches $180^\circ$ at $\vert\varphi\vert\sim 1.8^\circ$ or $\vert\varphi-60^\circ\vert\sim 2^\circ$. These two kinds of TBCIs are different in their interlayer exchange fields {  (defined as $E_i=\frac{9}{4}\sum_jJ_{ij}$)}. The AFM exchange field centers in the TBCI with $\varphi$ around $0^\circ$ form a kagome lattice (Figure \ref{fig2}c), while those around $60^\circ$ form a triagonal lattice (Figure \ref{fig2}g). Thus,  we name these TBCIs as K-TBCI and T-TBCI, respectively. Then the phase diagram of TBCI is composed of uniform FM phase, T-phase, and K-phase. 

In order to compare the two kinds of TBCIs in more detail, we studied a K-TBCI with $\varphi=1.7^\circ$ and a T-TBCI with $\varphi=61.7^\circ$. Their Mori\'{e} lattices have the same period length, with 4564 magnetic atoms in each period. Their interlayer exchange fields are shown in Figure \ref{fig2}c,g: there are three squeezed AFM islands in a period of panel (c), where the maximum AFM (FM) exchange field is 2.55 (-1.26) meV; and only one {  disc-shaped} AFM island in a period of panel (g), where the maximum AFM (FM) exchange field is 1.55 (-1.04) meV. The three AFM islands in panel (c) are isolated by narrow and shallow FM canals. The exchange fields along $A_1-A_2$ and $F_1-F_2$ are shown in Figure \ref{fig2}d. We can see that the FM canal is only about 30 \AA\, wide, and the barrier as shown by shaddow is lower than 0.3 meV/Cr. Thus, two neighboring AFM islands have chances to connect to each other, and the randomly occurring connections may lead to the disorder of magnetic states.  By contrast,  the AFM islands in panel (g)  are far away from each other. There are two typical paths (line $A_1-A_2$ and curve $A_1-A_3$) connecting two neighboring islands. The exchange fields (Figure \ref{fig2}h) along the paths show that there are very wide and deep barriers (the shaddows). The barrier widths are larger than 150 \AA\, and the barrier peaks along $A_1-A_3$ are about 1.0 meV/Cr, and those along $A_1-A_2$ are up to 1.5 meV/Cr. Therefore, two neighboring AFM islands are unlikely to fill up the barriers to form a connected area, and as a result, the system tends to have ordered magnetic domains.

We then simulated the steady magnetic states for the K-TBCI with $\varphi=1.7^\circ$ and the T-TBCI with $\varphi=61.7^\circ$. The simulations used $2\times2$ Mori\'{e} periods and random initial magnetic configurations. The results show that the AFM and FM areas are randomly distributed in the K-TBCI. Two typical magnetic states are illustrated in Figure \ref{fig2}e and \ref{fig2}f. On the other hand, the steady states for the T-TBCI are always isolated AFM domains surrounded by a connected FM domain as shown in Figure \ref{fig2}i,j. The simulations with other twist angles also show the same patterns: K-TBCIs have disorderly magnetic states, which agree well with the previous experiments\cite{TBCIexp1,TBCIexp2,TBCIexp3}; in contrast, the T-TBCIs have isolated AFM domains, which are in favor of skyrmions. Thus, even thou both K-phase and T-phase have co-existence of FM and AFM domains, their distributions are quite different.  Due to the ordered AFM domain distributions in T-TBCIs, we will focous on them in what follows. 

The detailed domain structure in T-TBCI varies for different twist angles as shown in Figure \ref{fig2}k-m. For $3.2^\circ<\vert\varphi-60^\circ\vert<7.3^\circ$, both the upper and lower layers have distorted magnetic spins, all Cr atoms in the island have positive $m_z$ components, and the in-plane components in different layers are in opposite directions.  But for $\vert\varphi-60^\circ\vert<3.2^\circ$, the two layers become unsymmetric. The upper layer has a skyrmion while the  lower layer has undistorted {  FM-order}. The skyrmion may also exist in the lower layer, and instead, the upper layer has undistorted {  FM-order}.

\subsection{Magnetic Skyrmions in T-TBCI}
Our further study shows that a variety of skyrmions may exist in T-TBCIs with $\vert\varphi-60^\circ\vert<3.2^\circ$. By setting many different initial spin configurations, we got different skyrmions. Examples are shown in Figure \ref{fig3}a, where the second one is a chiral Bloch-type Skyrmion. The $m_z$ components of skyrmions with three different twist angles are plotted in Figure \ref{fig3}b, which clearly show the spin flipping across the skyrmion edge. They have shapes of hyperbolic tangent functions, and thus the fitting lines are also plotted by using:

\begin{figure}
\centering
\includegraphics[width=.8\linewidth]{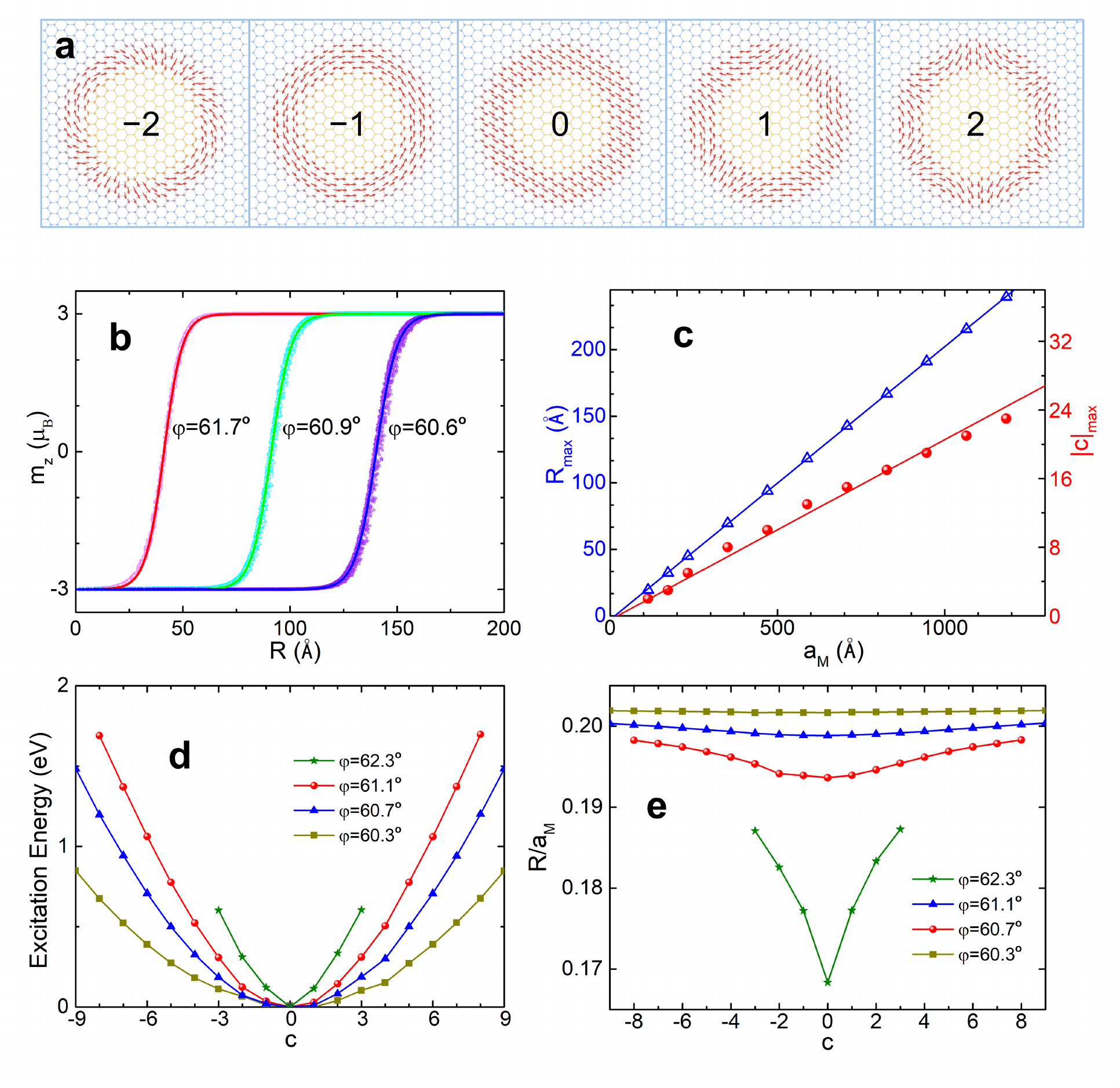}
\caption{(a) The skyrmions with c= -2, -1, 0, 1, and 2 in a T-TBCI with $\varphi=61.7^\circ$, the yellow and blue colors signs the magnetic moments along $-z$ and $z$ directions, respectively, and the red arrows at skyrmion edge show $m_x$ and $m_y$ components; (b) the $m_z$ component of Cr atoms as a function of the distance between the Cr atom and the skyrmion center, and the solid lines fitted with $m_0tanh((r-R)/w)$; (c) the maximum skyrmion charge and the associated skyrmion radius as functions of Mori\'{e} period length; (d,e) the excitation energy and skyrmion edge width as functions of skyrmion charge.}\label{fig3}
\end{figure}

\begin{equation}
\begin{aligned}
m_z(r,\phi)&=m_0tanh((r-R)/w),
\end{aligned}
\end{equation}  

where $R$ and $w$ are radius and width of the skyrmion, and the $m_0=3\,\mu_B$ is the saturated magnetic moment. Considering the direction of magnetic moment at skyrmion edge, the $m_x$ and $m_y$ components as functions of $r$ and polar angle $\phi$ can be written as: 

\begin{equation}
\begin{aligned}
m_x(r,\phi)&=m_0\frac{cos(\phi_0-c\phi)}{cosh((r-R)/w)}\\
m_y(r,\phi)&=m_0\frac{sin(\phi_0-c\phi)}{cosh((r-R)/w)}
\end{aligned}
\end{equation}

where $\phi_0$ is a phase factor. The number $c$ is a topological integer called skyrmion charge defined by 

\begin{equation}
\begin{aligned}
c=\frac{1}{4\pi}\int\mathbf{n}\cdot(\partial_x\mathbf{n}\times\partial_y\mathbf{n})\,\mathrm{d}x\mathrm{d}y,
\end{aligned}
\end{equation}

where $\mathbf{n}$ is the normalized magnetic moment $\mathbf{n}=\mathbf{m}/\vert\mathbf{m}\vert$. By using this equation, the calculated skyrmion charges for Figure \ref{fig3}a are -2, -1, 0, 1, and 2, respectively. The skyrmions with positive $c$ are also known as anti-skyrmions. 

The skyrmion radius $R$ increases as $\vert c\vert$ increases, as shown in Figure \ref{fig3}e, and both $R$ and the range of $c$ are dependent on $\varphi$. as shown in Figure \ref{fig3}b, a skyrmion with small value of $\vert\varphi-60^\circ\vert$ tends to have larger radius. The T-TBCI with $\varphi=62.5^\circ$ supports skyrmions with $c=-3\sim 3$, threfore the maximum skyrmion charge is $\vert c\vert_{max}=3$. All the skyrmions with $\vert c\vert >\vert c\vert_{max}$ are not stable during the LLG calculations. As shown in Figure \ref{fig3}c, the $\vert c\vert_{max}$ increases linearly with increasing of Mori\'e period length $a_M$, as $\varphi$ approaches $60^\circ$.

The excitation energy, defined as the energy difference between the system with a skyrmion and the system's ground state, is shown in Figure \ref{fig3}d. For a constant $c$, the excitation energy decreases with  $\varphi$ approaching $60^\circ$. For example, the excitation energy of skyrmion with $c=1$ in the T-TBCI with $\varphi=62.3^\circ$  is 115 meV, while that for $\varphi=60.3^\circ$ is only 8 meV. Since the skyrmion with $\varphi=60.3^\circ$ contains $\sim$18000 atoms, the 8 meV is a very small amount of energy. From Figure \ref{fig3}d, we can see that the trivial skyrmion ($c=0$) always has the lowest energy, and the excitation energy increases as $\vert c\vert$ increases.

\subsection{Potential Applications}
\begin{figure}
\centering
\includegraphics[width=.8\linewidth]{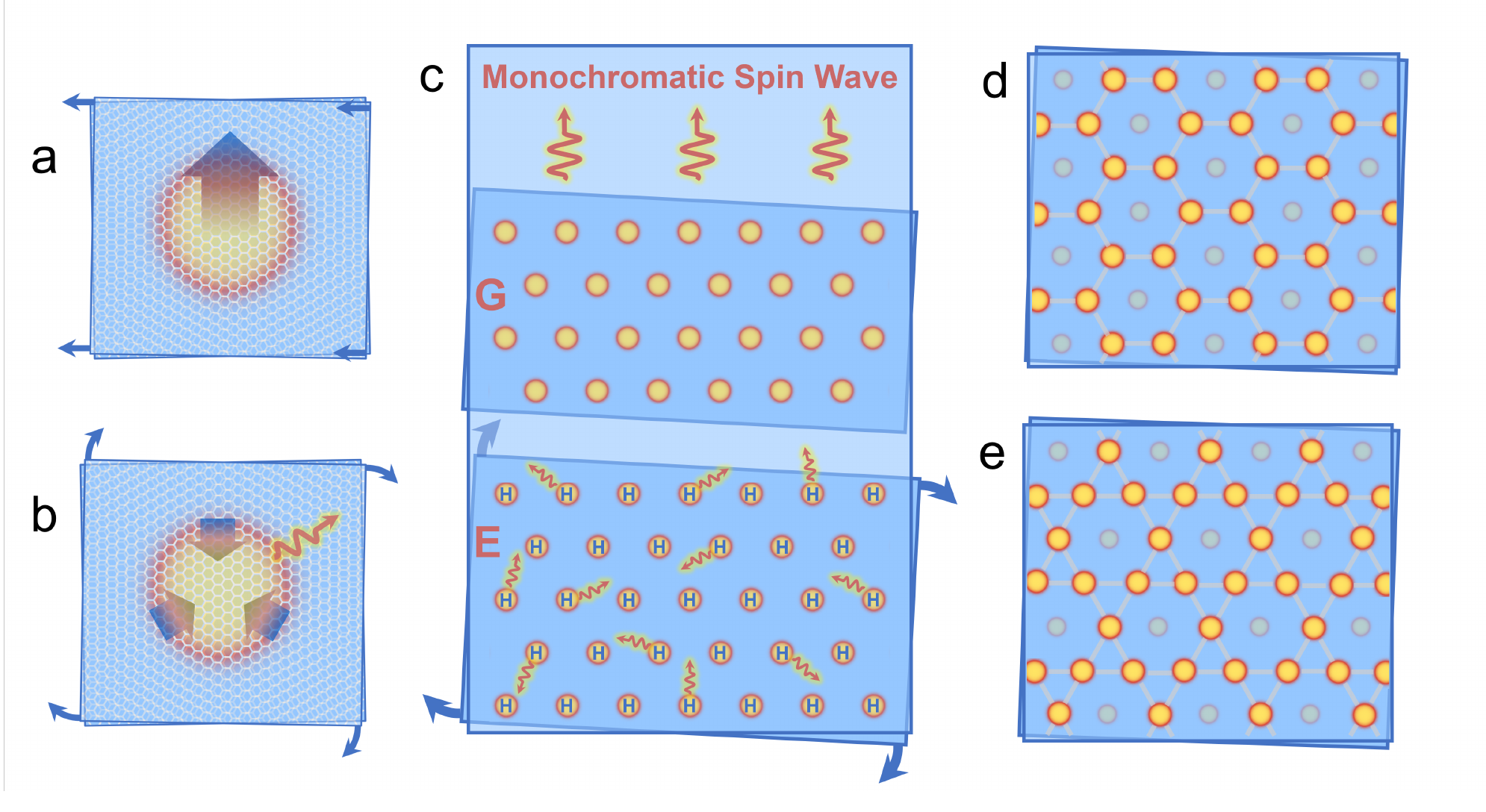}
\caption{(a) A large upward translation of skyrmion caused by a small left moving of a CrI$_3$ layer in a T-TBCI; (b) skyrmion compressed by changing twist angle, and the skyrmion collapse at the critical point; (c) the structure of a monochromatic spin wave generator based on T-TBCI, where the H on skyrmions shows the high skyrmion charges ; (d,e) the kagome and honeycomb skyrmion lattices (bright bubbles) at the upper layers and the triagonal skyrmion lattices (shadowy bubbles) at the lower layers.}\label{fig4}
\end{figure}

Due to the orderly distributed AFM domains and the existence of a variety of skyrmions, the T-TBCI can be used in device applications and fundamental research. Here we provide a few conceptual designs. Firstly, the T-TBCI with $\vert\varphi-60^\circ\vert<3.2^\circ$ contains a skyrmion lattice and can be used as a memory device. The informatoin can be coded into two channels. The skyrmion charge is an integer number with many options. It is topologically protected, thus  suitable to store information. Besides that, the skyrmion position can also be used to store information. As we have shown that a skyrmion in T-TBCI can stay either at the upper layer or at the lower layer and these two {  states} are degenerate. Therefore, the position of a skyrmion can store 1 bit information. It is well known that the skyrmions in magnetic alloy films\cite{Fert2017magnetic} are free to move, and their positions can be controlled by an electric current. By contrast, the skyrmions in T-TBCI are pinned at the AFM domain. Thus, in T-TBCI, the skyrmions with the stored information can not be transported by an electric current. Fortunately, the in-plane of one CrI$_3$ layer can be used to transport the skyrmions. As shown in Figure \ref{fig4}a, a small left move of the upper CrI$_3$ layer would lead to a large up move of the skyrmion. Actually, the motion of the skyrmion is always perpendicular to and much faster than that of the CrI$_3$ layer. The theoretical analysis can be found in the Supporting Information. 

Besides translation, the rotation can also be used to tune the skyrmion in T-TBCI. The rotation of one CrI$_3$ layer changes the twist angle $\varphi$, which alters the size of the Mori\'{e} pattern. Then, the size of AFM domain, i.e. the size of skyrmion, is altered accordingly. As shown in Figure \ref{fig4}b, the clockwise rotation of the upper CrI$_3$ layer makes $\varphi$ deviate from $60^\circ$, and compresses the skyrmion. A skyrmion with smaller radius has a smaller maximum skyrmion charge $\vert c\vert_{max}$, therefore the skyrmion charge $c$ may exceed $\vert c\vert_{max}$ during the compressing. Then, the skyrmion would collapse to another one with a lower $c$. The skyrmion would collapse completely if $\vert\varphi- 60^\circ\vert>3.2^\circ$, where the TBCI would have a uniform FM state. The collapse releases the redundent energy in many forms including spin waves. Based on that, the second application, a monochromatic spin wave generator was designed as shown in Figure \ref{fig4}c. It is composed of one large lower CrI$_3$ layer, and two smaller upper CrI$_3$ layers. The upper CrI$_3$ layer at emitter (E) region rotates slowly away from $60^\circ$. Then the skyrmions with high skyrmion charges at the lower CrI$_3$ layer will shrink their radiuses. The following collapses emit spin waves with wide range of frequencies along all in-plane directoins. After that, these spin waves will be filtered by the skyrmion lattices at the grating (G) region. The twist angle of the G region is kept fixed, then the skyrmions at the lower layer form a steady lattice, which only allows spin wave with certain frequencies to pass through.

In the foundamental research, the T-TBCI is a promissing platform for the study of spin topology and dynamics. For example, by using the degenerate upper and lower positions of skyrmions, a variaty of skyrmion lattices can be designed in T-TBCIs. As shown in Figure \ref{fig4}d,e, the skyrmions at the upper layers form honeycomb and kagome lattices respectively; and the both lower layers have sparse triagonal lattices. The rich skyrmion lattice types combined with the multiple choices of skyrmion charges make T-TBCI an intriguing platform for searching topologically nontrivial spin waves, especially when the system contains chiral Bloch-type skyrmions.

\section{Conclusion}
In summary, we have proposed the cluster TBCI models and SANNs, which enable us to calculate the functions of interlayer exchange interactions and efficiently predict these interactions in an arbitrarily stacked CrI$_3$ bilayer. Besides TBCI, these methods can also be used in the study of twisted CrI$_3$ multilayers and the stacking structures of other layered magnetic materials. The following LLG calculations predicted that T-TBCI has orderly distributed skyrmions, which can be used in novel device applications. Furthermore, a variety of skyrmion lattices can be designed in T-TBCI, where topologically nontrivial spin waves may be found. Due to these unique properties, we call on further theoretical and experimental studies on this new material.

\section{Methods}

\subsection{Artificial Neural Network}
We have constructed two SANNs in Matlab\cite{MATLAB:R2021a} to predict $J$ for any $r$, $\theta$, $z$, and $\varphi$ values. As we have discussed, $J_{BB}$ and $J_{BA}$ can be directly obtained from $J_{AA}$ and $J_{AB}$, therefore we only need to predict the values of $J_{AA}$ and $J_{AB}$. Due to their different symmetry properties, we used two SANNs for $J_{AA}$ and $J_{AB}$ separately. The two SANNs have the same structure as shown in Figure \ref{fig1}b. There are 18 weight shared feed forward neural networks (FFNNs) in a SANN. Each FFNN contains four hidden layers with $40\times100\times100\times40$ neurons. We have also checked the FFNNs that have three hidden layers with $20\times60\times20$, $30\times80\times30$, $40\times100\times40$ neurons, and four hidden layers with $20\times60\times60\times20$, $30\times80\times80\times30$. After training and testing them on a {  GeForce RTX} 3080Ti GPU, the FFNN with $40\times100\times100\times40$ neurons shows the best prediction accuracy.  

There are five input variables: $x=rcos\theta$, $y=rsin\theta$, $z$, $cos\varphi$, and $sin\varphi$. The use of $cos\varphi$ and $sin\varphi$ rather than $\varphi$ naturally assures the periodicity of $J$ on $\varphi$. We used 5600 DFT+MFT data points for each SANN. The number of data points are extended to 100,800 according to the symmetry operations. And we randomly choose 80\% of them to be the training dataset, and the remaining 20\% data points to be the testing dataset. After training $1\times10^5$ scaled conjugate gradient (SCG) cycles, the neural networks could predict $J$ values in high accuracy. The regression accuracies for $J_{AA}$ and $J_{AB}$ are 99.98\% and 99.95\% with squred mean error 0.007 meV and 0.011 meV, respectively.  In the real applications, we have adopted the same settings and used all the DFT+MFT data point as the training dataset. 

\subsection{Magnetic Structure Calculation}

After obtained the interlayer parameters from the SANNs and intralayer parameters from DFT calculations, we then calculated magnetic properties of the system. The
equation of motion describing such a classical spin model is Landau-Lifshitz-Gillbert (LLG) equation. 

\begin{equation}
\begin{aligned}
\frac{d\mathbf{S}_i}{dt}=\gamma\mathbf{S}_i\times\frac{\partial H}{\partial \mathbf{S}_i}+\gamma\alpha\mathbf{S}_i\times(\mathbf{S}_i\times\frac{\partial H}{\partial \mathbf{S}_i}),
\end{aligned}
\end{equation}

where $\gamma$ is the gyromagnetic ratio, and $\alpha$ is the Gilbert damping coefficient, whose positive value (0.5$\sim$1.0 in our calculations) ensures that the system converges to a steady magnetic state. The $-\frac{\partial H}{\partial \mathbf{S}_i}$ in the last term works as an effective magnetic field. 

The steady magnetic structures for K-TBCI and T-TBCI were obtained by random initial magnetic configurations. In order to exaust all different skyrmions, we also used variety of  vortex-shaped initial magnetic configurations.

\medskip
\textbf{Acknowledgements} \par 
The author would like to thank Dr. Chen Si, Associate Professor of Beihang University, for the disccusion on the neural networks, Dr. Jize Zhao, Professor of Lanzhou University, for the discussion on spin lattice models of CrI$_3$ layer, and Dr. Yue Ji, Assistant Professor of Beijing Insitute of Technology, for language editing. This work was financially supported by National Natural Science Foundation of China (Grants Nos. 12022415 and 11974056).


\end{document}


\section{Symmetry of $J$-Functions}
There are four kinds of interlayer $J$-functions for TBCI, they are $J_{AA}$, $J_{AB}$, $J_{BA}$, and $J_{BB}$. Each function can be written in the form $J=J(r,\theta,z,\varphi)$, where $(rcos\theta,rsin\theta,z)=\mathbf{r}_2-\mathbf{r}_1$ is the relative positions of the two Cr atoms, and $\varphi$ is the twist angle. They are constrained by the symmetry of CrI$_3$ single layer and TBCI. The point group of a CrI$_3$ single layer is D$_{3d}$, which has a $C_3$ symmetry operation rotating around a Cr atom or a hollow site center. Suppose we have a twisted bilayer CrI$_3$. Lets consider a Cr atom at $(0,0,-d/2)$ in unrotated lower layer and another Cr atom at $(rcos\theta,rsin\theta,d/2)$ in the $\varphi$-rotatted top layer.  Considering the $C_3$ symmetry operations around Cr atoms, the function $J(r,\theta,z,\varphi)$ has the following properties 

\begin{figure}
\centering
\includegraphics[width=11.4 cm]{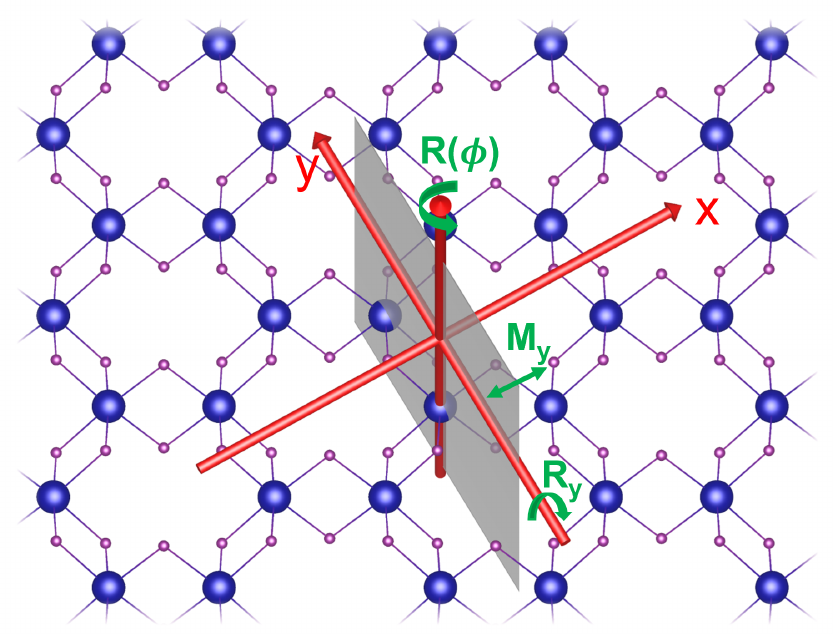}
\caption{\label{figS1} The coordinate system and symmetry operations of a TBCI. Only the lower CrI$_3$ layer is shown here to have a better view, and the coordinate origin is at the middle of the two CrI$_3$ layers. }
\end{figure}

\begin{equation}
\begin{aligned}
J(r,\theta,z,\varphi)&=J(r,\theta\pm\frac{2\pi}{3},z,\varphi)\\
&=J(r,\theta,z,\varphi\pm\frac{2\pi}{3}).
\end{aligned}
\end{equation}

In order to discuss the other symmetry operations conviniently, we name a Cr atom in the sublattice-A of the lower layer as $A(r,\theta,-d/2,\varphi_1)$, where the first three variables are the cylindrical coordinates, and $\varphi_1$ is the rotation angle of the lower layer. Similarly, we define the three other types of Cr atom as $B(r,\theta,-d/2,\varphi_1)$,  $A(r,\theta,d/2,\varphi_2)$, and $B(r,\theta,d/2,\varphi_2)$. The two Cr layers are seperated by distance $z=d$, and the twist angle of the bilayer is $\varphi=\varphi_2-\varphi_1$.  Lets consider the rotation $R(\phi)$ as shown in Figure S1, which is the $\phi$-rotation around axis $r=0$. Then it follows

\begin{equation}
\begin{aligned}
R(\phi)A(r,\theta,z,\varphi)&=A(r,\theta+\phi,z,\varphi+\phi)\\
R(\phi)B(r,\theta,z,\varphi)&=B(r,\theta+\phi,z,\varphi+\phi). 
\end{aligned}
\end{equation}

Another useful operation is  $R_y$, which is the $\pi$-rotation around $y$ axis ($z=0,\theta=\pi/2$). Then we have

\begin{equation}
\begin{aligned}
R_yA(r,\theta,z,\varphi)&=A(r,\pi-\theta,-z,-\varphi) \\
R_yB(r,\theta,z,\varphi)&=B(r,\pi-\theta,-z,-\varphi) 
\end{aligned}
\end{equation}

The third operation is $T(r,\phi)$, which is the in-plane translation with vector $(rcos\phi,rsin\phi)$. Lets only consider the folowing four special cases

\begin{equation}
\begin{aligned}
T(r,\phi)A(0,0,z,\varphi)&=A(r,\phi,z,\varphi)\\
T(r,\phi+\pi)A(r,\phi,z,\varphi)&=A(0,0,z,\varphi)\\
T(r,\phi)B(0,0,z,\varphi)&=B(r,\phi,z,\varphi)\\
T(r,\phi+\pi)B(r,\phi,z,\varphi)&=B(0,0,z,\varphi)
\end{aligned}
\end{equation}

The last operation is $M_y$, the mirror operation with respect to plain $y=0$. It has the following effect

\begin{equation}
\begin{aligned}
M_yA(r,\theta,z,\varphi)&=B(r,-\theta,z,-\varphi)\\
M_yB(r,\theta,z,\varphi)&=A(r,-\theta,z,-\varphi)
\end{aligned}
\end{equation}

Considering a Cr atom at the lower layer and another Cr atom at the upper layer, the combined using of these operations to a TBCI will be

\begin{equation}
\begin{aligned}
M_y&[B(0,0,-d/2,0)B(r,\theta,d/2,\varphi)] \\
&= A(0,0,-d/2,0)A(r,-\theta,d/2,-\varphi)  \\
M_y&[B(0,0,-d/2,0)A(r,\theta,d/2,\varphi)] \\
&= A(0,0,-d/2,0)B(r,-\theta,d/2,-\varphi)  \\
R(\varphi&)T(r,-\theta)R_y[A(0,0,-d/2,0)A(r,\theta,d/2,\varphi)]\\
&=R(\varphi)T(r,-\theta)[A(r,\pi-\theta,-d/2,-\varphi)A(0,0,d/2,0)]    \\
&=R(\varphi)[A(0,0,-d/2,-\varphi)A(r,-\theta,d/2,0)] \\
&=A(0,0,-d/2,0)A(r,-\theta+\varphi,d/2,\varphi)\\
M_y&R(\varphi)T(r,-\theta)R_y[A(0,0,-d/2,0)B(r,\theta,d/2,\varphi)]\\
&=M_yR(\varphi)T(r,-\theta)[B(r,\pi-\theta,-d/2,-\varphi)A(0,0,d/2,0)]    \\
&=M_yR(\varphi)[B(0,0,-d/2,-\varphi)A(r,-\theta,d/2,0)] \\
&=M_y[B(0,0,-d/2,0)A(r,-\theta+\varphi,d/2,\varphi) ] \\
&=A(0,0,-d/2,0)B(r,\theta-\varphi,d/2,-\varphi).
\end{aligned}
\end{equation}

Therefore, the $J$-functions between the two Cr atoms have the following properties.

\begin{equation}
\begin{aligned}
J_{BB}(r,\theta,z,\varphi)&=J_{AA}(r,-\theta,z,-\varphi) \\
J_{BA}(r,\theta,z,\varphi)&=J_{AB}(r,-\theta,z,-\varphi) \\
J_{AA}(r,\theta,z,\varphi)&=J_{AA}(r,-\theta+\varphi,z,\varphi) \\
J_{AB}(r,\theta,z,\varphi)&=J_{AB}(r, \theta-\varphi,z,-\varphi) \\
\end{aligned}
\end{equation}

The first two equations show that there are only two independent $J$-functions, and the last two equations further simplifies the two independent $J$-functions. 

\section{Size of the Cluster Model}
We have used a cluster model to calculate the functions of interlayer exchange interaction. The model contains two parallel pieces of CrI$_3$, each piece contains 13 Cr, 48 I, and 9 Li atoms. As shown in Figure S2a,b, the central Cr atoms are in sublattice A and B, respectively. The homo- and hetero bilayer clusters composed of the two pieces can be used to calculate $J_{AA}$ and $J_{AB}$, respecitively. The $J$ functions are obtained by performing the magnetic force theory (MFT) calculations after the DFT charge convergence. The DFT and MFT calculations on cluster models are performed by using OpenMX package\cite{openmx}, which is based on norm-conserving pseudopotentials\cite{NCPP1,NCPP2,NCPP3,NCPP4,NCPP5} and pseudo-atomic localized basis functions\cite{AtomicOrbital1,AtomicOrbital2}. We have used generalized gradient approximation (GGA) in the Perdew, Burke, and Ernzerhof (PBE) form\cite{PBE} to the exchange-correlation functional. The weak interlayer vdW interactions were described by using DFT-D3 method\cite{vdW}. The repulsion between $d$-electrons on Cr atoms are described by a Hubbard correction with 3 eV on-site Coulomb repulsion\cite{DFTU}. 

\begin{figure}
\centering
\includegraphics[width=11.4 cm]{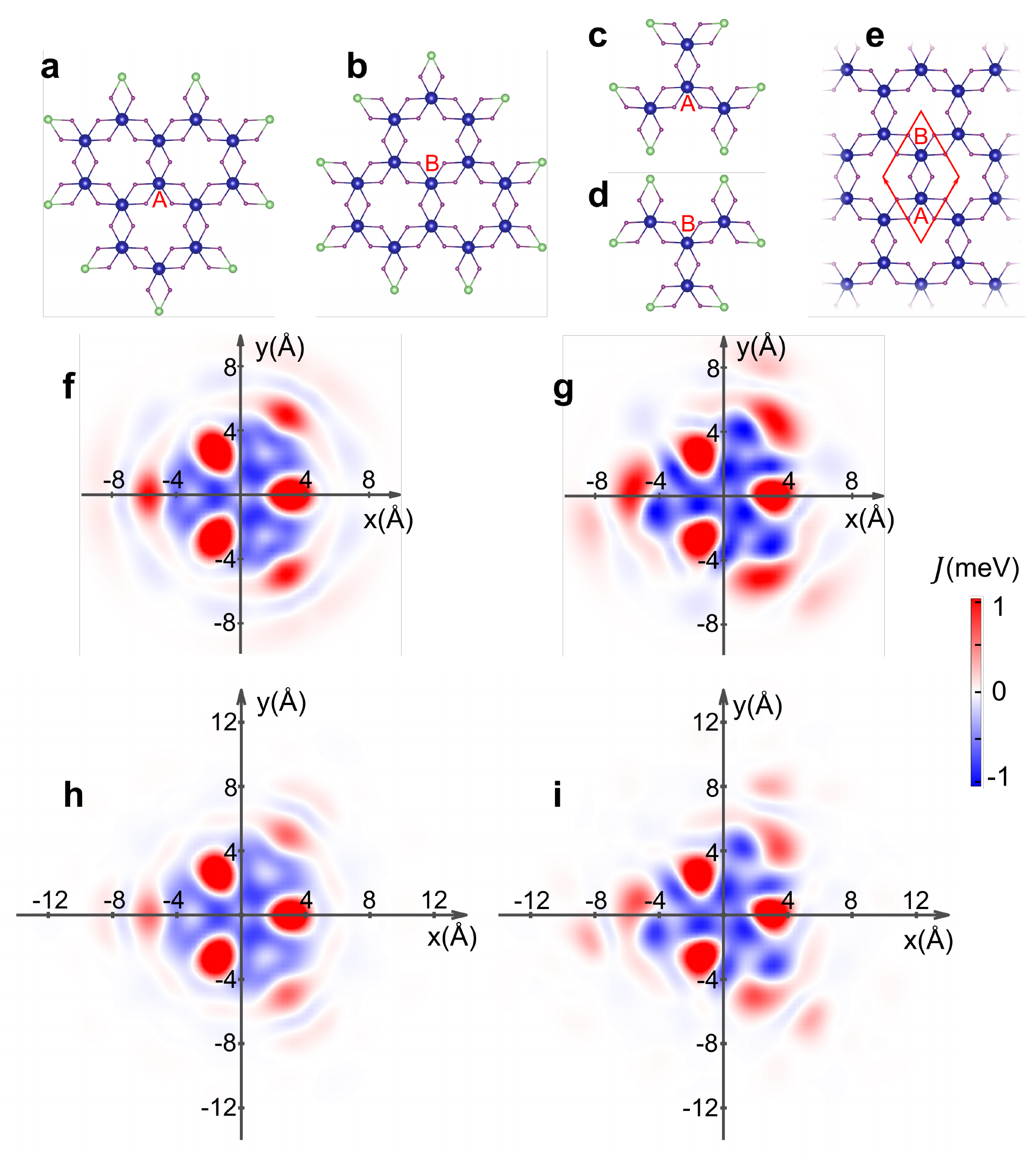}
\caption{\label{figS2}
(a,b) The large CrI$_3$ clusters; (c,d) the small CrI$_3$ clusters; (e) the periodic CrI$_3$ layer. The 'A' and 'B' are used to show that the Cr atoms are in sublattice-A and sublattice-B, respectively. (f,g) The values of $J_{AA}$ and $J_{AB}$ at $\varphi=0^o$ calculated using the small clusters; (h,i) The values of $J_{AA}$ and $J_{AB}$ at $\varphi=0^o$ calculated using the periodic CrI$_3$.
}
\end{figure}

The size of each CrI$_3$ piece should be large enough, otherwise the interlayer magnetic exchange interaction may be different with that of a periodic TBCI.  The radius of our model is 11.9 \AA. Actually, we have tested a model with smaller radius as shown in Figure S2c,d. The calculated results are shown in Figure S2f,g. The $J$-functions at $\varphi=0^o$ with two Cr atoms at arbitrary distance can be obtained in a periodic model (Figure S2e), as shown in Figure S2h,i, which offers the standard to check the accuracy of cluster models.  

We can see that the Figure S2f,g agree with Figure S2h,i for short distances ($r<$4 \AA), while they are different for long distances (see the red areas at $r>$4 \AA). Thus, the small cluster model is only valid for $r<4$ \AA.  On the other hand, the $J$-functions obtained using the large cluster model as shown in Figure 1g,i agree well with Figure S2h,i up to $r<9$ \AA, and the values beyond $9$ \AA\, are omissible. Therefore, our large cluster model correctly describes the interlayer magnetic exchange interactions in TBCI.

\section{Edge Structure of the Cluster Model}
In a periodic CrI$_3$ layer, each Cr atom losts 3 electrons, and each I atom receives 1 electron. However, this balance can be easily destroyed at the edge of a cluster model. Therefore, the edge should be handled carefully to keep the charge balance. We have tested two different edge settings as shown in Figure S2a,b and Figure S3a,b. The cluster model was cutted from a periodic CrI$_3$ layer. The Cr-I bonds at the boundary were cutted with outer side Cr atoms deleted. Thus, the cluster model has surplus I atoms at the edge. In the first edge setting, we provided extra Li atoms at the deleted Cr sites. The radius of a Li atom is 1.5 \AA, wihch is close to that of a Cr atom (1.3 \AA). And, each Li atom provides 1 electrons to edge I atoms, keeping the -1 oxidation state of I. In the second edge setting, we deleted half the number of edge I atoms to keep the oxidation state of I. However the rotation and reflection symmetries of the system is destroyed. In order to restore these symmetries, we shifted the remainder edge I atoms to the symmetric positions.

\begin{figure}
\centering
\includegraphics[width=11.4 cm]{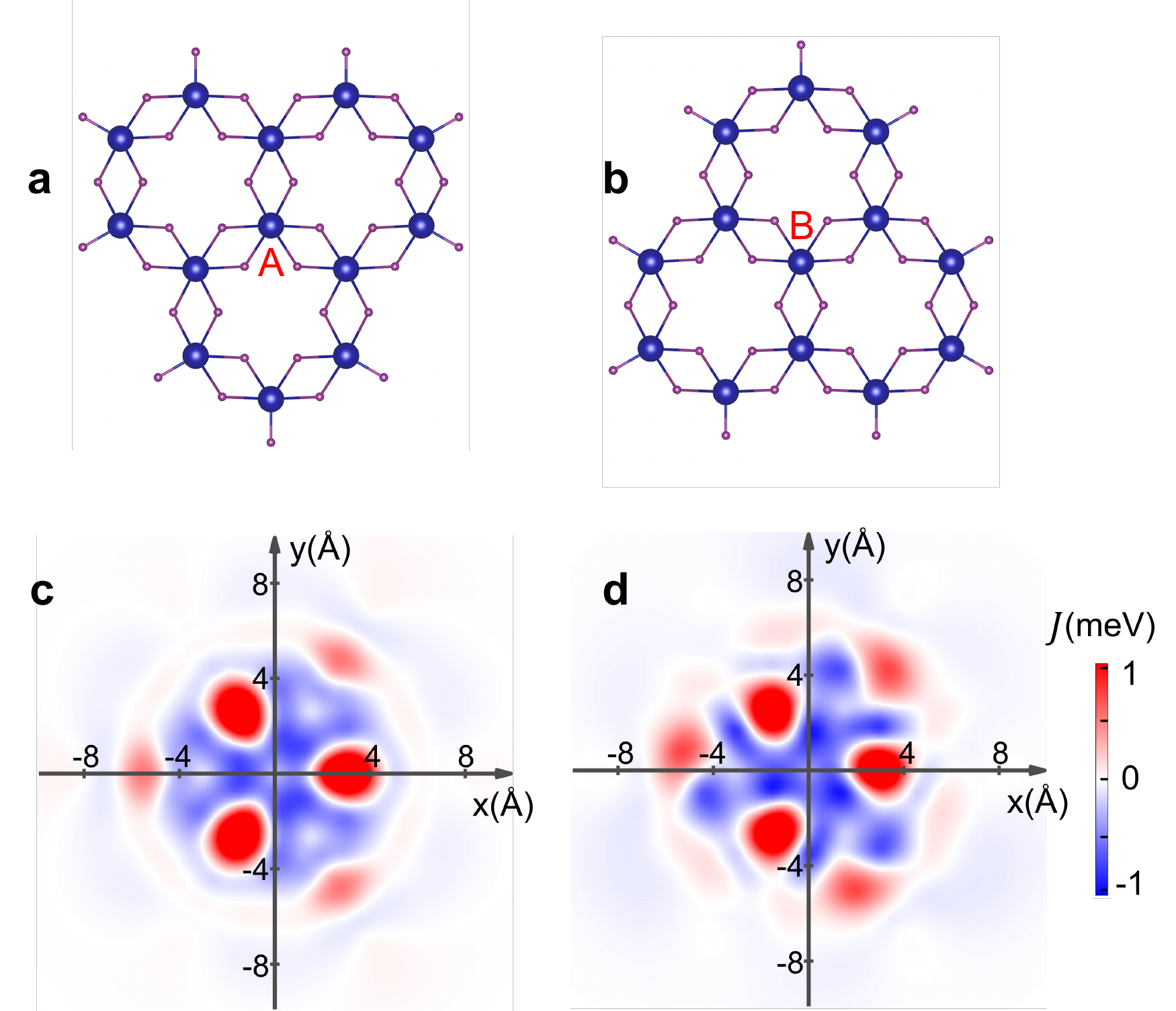}
\caption{\label{figS3}
(a,b) The CrI$_3$ pieces with half the number of edge I atoms. (c,d) The values of $J_{AA}$  and $J_{AB}$ at $\varphi=0^o$ calculated using TBCI cluster models with half the number of edge I atoms.
}
\end{figure}

As we have shown in the last section, the first edge setting works well. Here, we show the results of the second edge setting in Figure S3c,d. We can see that the $J_{AA}$  for $r<$7 \AA\, and $J_{AB}$ for $r<$4 \AA\, agree well with Figure S2h,i. While the $J_{AB}$ for $r>$4 \AA\, are noticeably different with  Figure S2i, it is even more worse than that of the small cluster model with extra Li atoms (Figure S2g).  This is because the deleting of edge I atoms altered the crystal field at the edge Cr atoms, then largely changed the $J$ values. Thus the cluster model with extra Li atoms is a better choice.

\section{Interlayer Exchange Energy in Bilayer CrI$_3$}
As shown in Figure 3b, the FM-AFM coexistence state emerges when the twist angle approaches $0^o$ or $60^o$. For the other twist angles, the most stable states are FM. As studied previously\cite{Sivadas,Yu}, the interlayer interaction energy between two FM CrI$_3$ layers can be effectively tuned by in-plain translation. The SANNs in this work enable us to predict interlayer exchange interaction energy between two FM CrI$_3$ layers with any twist angle. Here, we predict the interlayer exchange interaction energies for $\varphi=0^o$ and $60^o$. The results for rigid shift are shown in Figure S4a,b. The results in panel (a) agrees well with Ref\cite{Sivadas}. The interlayer distance may be different for different translations, therefore we adopted a dipole force intropolated interlayer distances.  The relationship between interlayer distance and $z$-component of electric dipole interaction was fitted in a polynomial. Then it was used to model the local interlayer distances in large TBCI systems. Based on this scheme, the interlayer exchange energies are shown in Figure S4d,e. The interlayer exchange energies for $\varphi=60^o$ with relaxed structures are already calculated in Ref\cite{Yu}. We can see that the Figure S4e agrees well with the DFT results.

\begin{figure}
\centering
\includegraphics[width=11.4 cm]{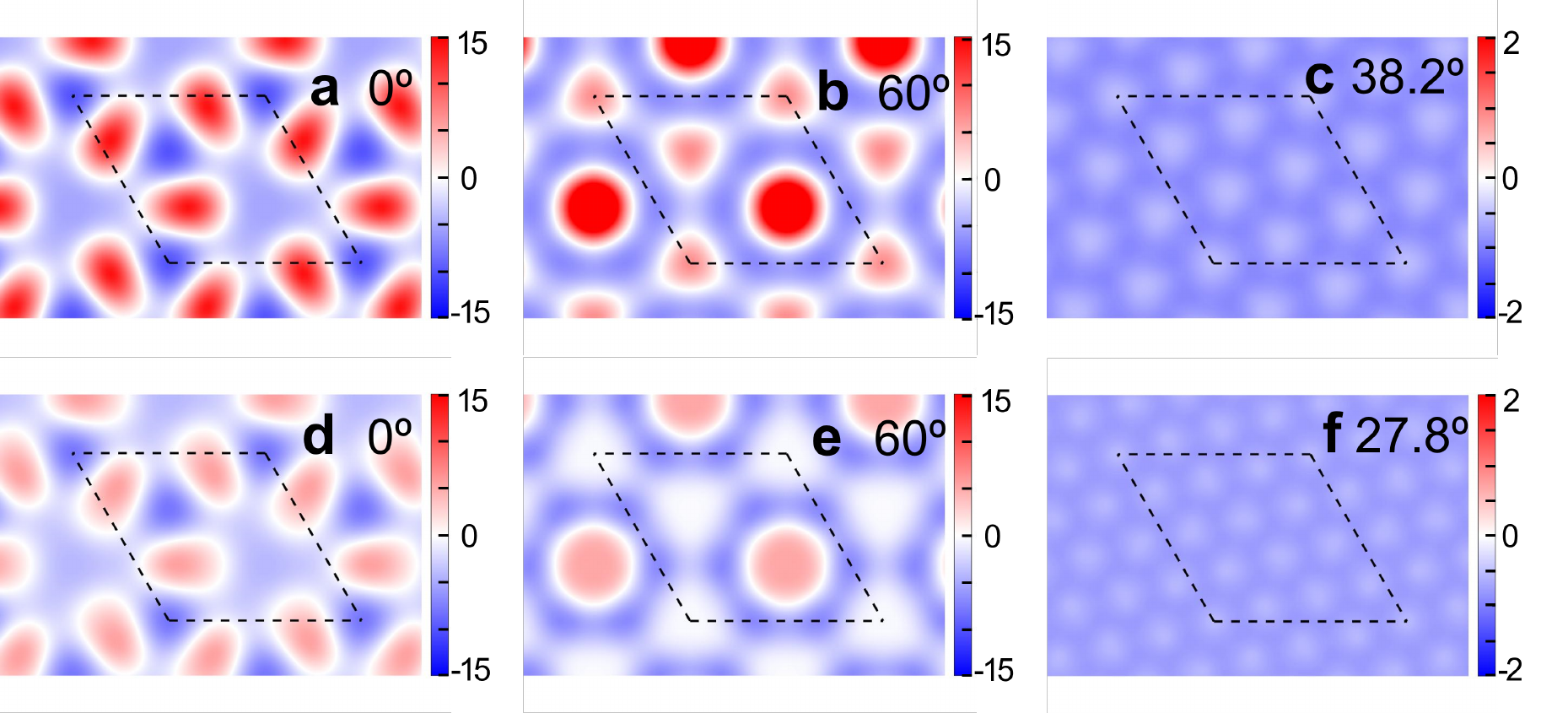}
\caption{\label{figS4}
The interlayer exchange energy for TBCI at FM state with $\varphi=0^o$ (a,d), $60^o$ (b,e), $38.2^o$ (c), and $27.8^o$ (f), respectively. The panel (a) and (b) shows the results for rigid translation with 6.62 \AA\, interlayer distance, while the other panels are the results for TBCIs with dipole force intropolated interlayer distances. The dashed  rhombus in these figures show a unit cell.}
\end{figure}

The calculation results for two other twist angles are also shown in Figure S4c,f. We can see that the interlayer exchange energies have more peaks than that for  $\varphi=0^o$ and $60^o$, and the peaks are much weaker.  That is because there are much more magnetic atoms, the positive and negative interlayer exchange interactions have more chances to cancel with each other.

\section{Four-State Method}
The intralayer magnetic exchange interactions are calculated directly from DFT by using the four-state-method\cite{xiang2013magnetic,PhysRevB.84.224429,ChangsongXu}. The magnetic Hamiltonian in a CrI$_3$ layer can be written as

\begin{equation}
\begin{aligned}
H=\sum_{i>j}\mathbf{S}_i^T\mathcal{J}_{ij}\mathbf{S}_j
\end{aligned}
\end{equation}

where $\mathbf{S}_i$ is the spin of the $i$-th Cr atom, and $\mathcal{J}_{ij}$ is the full $3\times 3$ exchange interaction matrix between the $i$-th and $j$-th Cr atoms. 

In order to calculate the magnetic interaction between site-$m$ and site-$n$, we write the Hamiltonian as: 

\begin{equation}
\begin{aligned}
H     &= H_{mn} + H_{mo} + H_{no} + H_{o} \\
H_{mn}&= \mathbf{S}_m^T\mathcal{J}_{mn}\mathbf{S}_n \\
H_{mo}&= \sum_{i\ne m,n}\mathbf{S}_m^T\mathcal{J}_{mi}\mathbf{S}_i \\
H_{no}&= \sum_{i\ne m,n}\mathbf{S}_n^T\mathcal{J}_{ni}\mathbf{S}_i \\
H_{o} &= \frac{1}{2}\sum_{i,j \ne m,n}\mathbf{S}_i^T\mathcal{J}_{ij}\mathbf{S}_j \\
\end{aligned}
\end{equation}

where the $H_{mn}$ is the interaction between site-$m$ and site-$n$, the $H_{mo}$ and $H_{no}$ describe the interactions of site-$m$ and site-$n$  with the other sites, and all the other interactions are in $H_o$.

Then the matrix $\mathcal{J}_{mn}$ can be obtained by the energies of four different magnetic states. In these states, the spins of $m$- and $n$-th Cr atoms are $(\mathbf{S}_m, \mathbf{S}_n)$, $(-\mathbf{S}_m, \mathbf{S}_n)$, $(\mathbf{S}_m, -\mathbf{S}_n)$, and $(-\mathbf{S}_m, -\mathbf{S}_n)$, respectively.  All the spin for the other Cr atoms are unchanged in the four magnetic states. The associated total energies are named as $E_1$, $E_2$, $E_3$, and $E_4$, respectively. Then the term $H_{mn}$ can be isolated as 

\begin{equation}
\begin{aligned}
\mathbf{S}_m^T\mathcal{J}_{mn}\mathbf{S}_n=\frac{E_1-E_2-E_3+E_4}{4}.
\end{aligned}
\end{equation}

All the nine matrix elements of $\mathcal{J}_{mn}$ can be obtained by approprate settings of $\mathbf{S}_m$ and $\mathbf{S}_n$. In our DFT calculations, the other sites were setted in a direction perpenticular to both $\mathbf{S}_m$ and $\mathbf{S}_n$ to supress the influence of higher order interactions. For example, in the calculation of $\mathcal{J}_{xy}$, we set $\mathbf{S}_m=(3/2,0,0)$, $\mathbf{S}_n=(0,3/2,0)$,  and the spin of all the other sites as $(0,0,3/2)$.

\section{Symmetry of the Intralayer Magnetic Interaction}
By using the four-state method, we have calculated the exchange interactions between Cr atoms upto the third nearest neighbors, as shown in Figure S5. There are 3, 6, and 3 neighbors for the first, second, and third nearest neighbors, respectively. Since the system has non-negligible spin-orbit couplings, the $3\times3$ exchange matrices for the first nearest neighbors are neither diagonal nor identical. They are related by matrix $R_z$, which describes a $2\pi/3$-rotation with respect to $z$-axis crossing the central Cr atom. 

\begin{figure}
\centering
\includegraphics[width=11.4 cm]{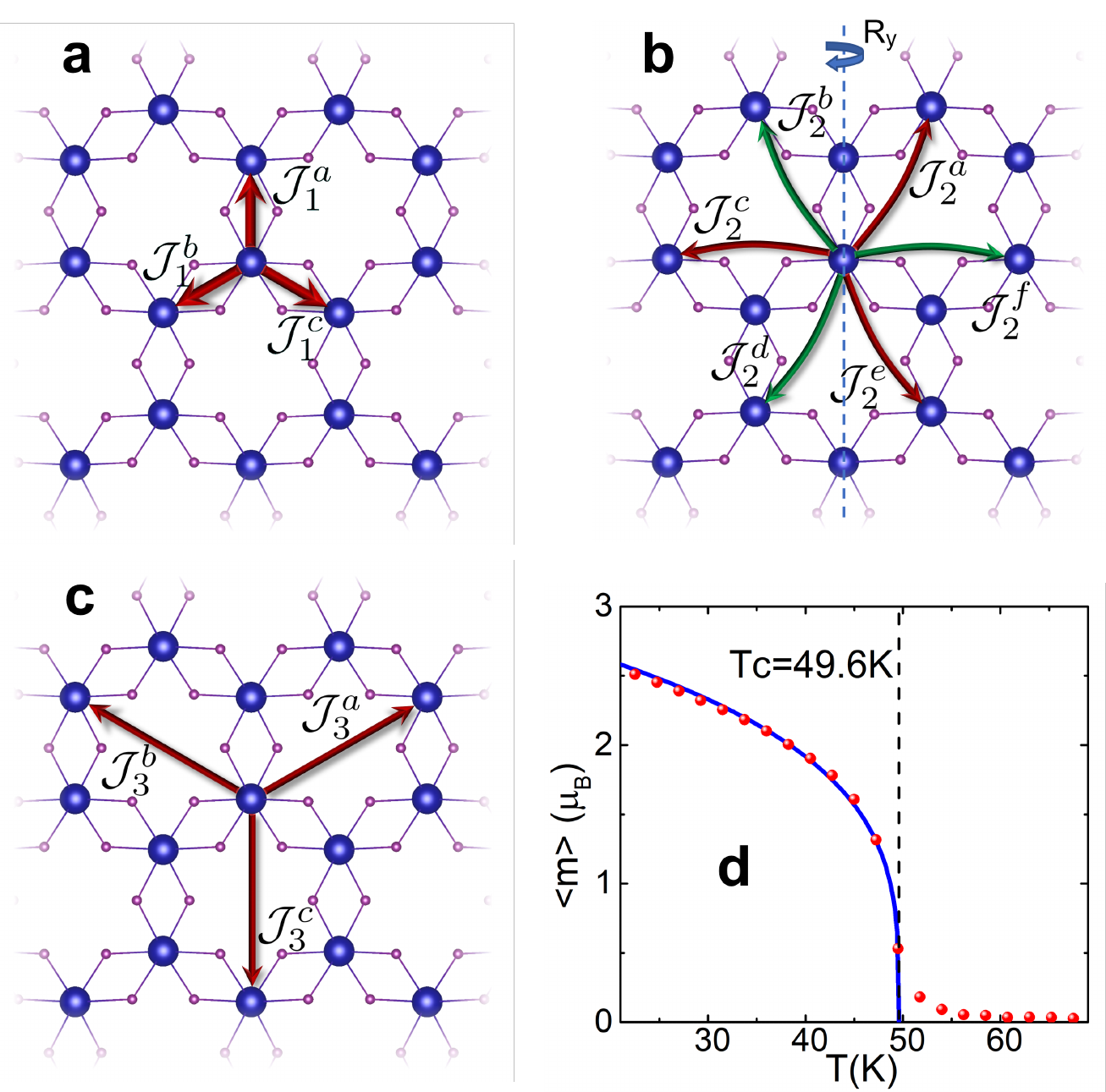}
\caption{\label{figS5}
 The exchange interactions for (a) the first, (b) second, and (c) third nearest neighbors. The red and green interactions in panel (b) are related by the rotation $R_y$. (d) The magnetization of a single layer CrI$_3$ as a function of temperature, simulated by using Monte Carlo method. 
}
\end{figure}

\begin{equation}
\begin{aligned}
\mathcal{J}_1^b=R_z\mathcal{J}_1^aR_z^{-1}\\
\mathcal{J}_1^c=R_z\mathcal{J}_1^bR_z^{-1}
\end{aligned}
\end{equation}

The three magnetic exchange matrices for the third nearest neighboring Cr atoms are also related by $R_z$.

\begin{equation}
\begin{aligned}
\mathcal{J}_3^b=R_z\mathcal{J}_3^aR_z^{-1}\\
\mathcal{J}_3^c=R_z\mathcal{J}_3^bR_z^{-1}
\end{aligned}
\end{equation}

The magnetic exchange matrices for the second nearest neighboring Cr atoms can be devided into two groups, as shown by the red and green arrows in Figure S5b. The matrices in the same group are related by $R_z$.

\begin{equation}
\begin{aligned}
\mathcal{J}_2^c=R_z\mathcal{J}_2^aR_z^{-1}\\
\mathcal{J}_2^e=R_z\mathcal{J}_2^cR_z^{-1}\\
\mathcal{J}_2^d=R_z\mathcal{J}_2^bR_z^{-1}\\
\mathcal{J}_2^f=R_z\mathcal{J}_2^dR_z^{-1}
\end{aligned}
\end{equation}

And the matrices in red group can be converted to that in green group by matrix $R_y$, which describes the $\pi$ rotation around $y$-axis. Its inversion matrix is itself $R_y^{-1}=R_y$.

\begin{equation}
\begin{aligned}
\mathcal{J}_2^b=R_y\mathcal{J}_2^aR_y\\
\mathcal{J}_2^f=R_y\mathcal{J}_2^cR_y\\
\mathcal{J}_2^d=R_y\mathcal{J}_2^eR_y
\end{aligned}
\end{equation}

In this work, we calculated $J_1^a$, $J_2^a$, and $J_3^a$ based on DFT method by using Vienna Ab initio Simulation Package (VASP)\cite{vasp}.  The core electrons were described by PAW potentials\cite{PAW1,PAW2}. The exchange-correlation functional is a kind of meta generalized gradient approximation r$^2$SCAN\cite{R2SCAN}, which  matches the accuracy of the parent SCAN functional\cite{SCAN} but with significantly improved numerical efficiency and accuracy under low-cost computational settings. The cutoff energy for plane wave basis is 227 eV. We have used a unit cell slab model with 15 \AA\, vaccum space to relax the atoms until the force on each atom is smaller than 0.001 eV/\AA. The reciprocal space integration was performed in a 8$\times$8$\times$1 $\Gamma$-centered K-smesh. After relaxation, we constructed a  3$\times$3$\times$1 supercell to calculate the total energies for the four-state method.  In the supercell DFT calculations, the reciprocal space integration was performed in a 2$\times$2$\times$1 $\Gamma$-centered K-smesh.  The large supercell permits the calculation of magnetic exchange parameters upto the third nearest neighbors. The calculated results are

\begin{equation}
\begin{aligned}
\mathcal{J}_1^a=&\begin{bmatrix}
 -1.746 & \ \   0     &  -0.334 \\
 \ \  0     &  -2.430 &  \ \  0   \\
 -0.334 &  \ \  0     &  -2.289
\end{bmatrix}\\
\mathcal{J}_2^a=&\begin{bmatrix}
-0.598  & \ \  0.058  &  \ \  0.067 \\
-0.004  & -0.563  &  \ \  0.001 \\
-0.036  & \ \  0.029  & -0.543 
\end{bmatrix}\\
\mathcal{J}_3^a=&\begin{bmatrix}
  \ \ 0.154  & \ \   0     & -0.008  \\
   \ \    0  &  \ \  0.143 & \ \  0      \\
 -0.008  & \ \   0     & \ \  0.147
\end{bmatrix}
\end{aligned}
\end{equation}

The other exchange interaction matrices were obtained by using the rotation matrices. We have checked that the matrices $\mathcal{J}_1^b$, $\mathcal{J}_2^b$, and $\mathcal{J}_3^b$ converted from $\mathcal{J}_1^a$, $\mathcal{J}_2^a$, and $\mathcal{J}_3^a$ agree well with those directly obtained from DFT calculations.

In order to test these $\mathcal{J}$-matrices, we performed Monte Carlo simulations of a large supercell of CrI$_3$ single layer. There are 1862 Cr atoms in one supercell, simulated with 3$\times 10^7$ Monte Carlo steps. The magnetization was sampled in the last 1$\times 10^7$ steps. The simulation results are shown in Figure S5d. The Curie temperature ($T_c$) were obtained by fitting the formula $\left<m\right>=m_0(1-T/T_c)^\beta$, where $m_0=3\,\mu_B$. The fitting results show that $T_c$=49.6 K, which closes to the experimental value 45 K\cite{HuangXu2017}.

\section{Motion of Skyrmions}
In a TBCI, the positions of skyrmions depend on the stacking structures of the two CrI$_3$ layers. Therefore, the translation of one CrI$_3$ layer will also translate the positions of skyrmions. Lets consider a bilayer CrI$_3$ without twisting, both layers have the same lattice vectors $\mathbf{A}_1$ and $\mathbf{A}_2$, which are written in column vectors. We rotate the upper layer with a small angle $\varphi$ ($\ll\pi$), and translate in $xy-$plain with vector $\mathbf{t}$. Then the point $u_1\mathbf{A}_1+u_2\mathbf{A}_2$ in the upper layer will change to $u_1\mathcal{R}(\varphi)\mathbf{A}_1+u_2\mathcal{R}(\varphi)\mathbf{A}_2+\mathbf{t}=\mathcal{R}(\varphi)\mathcal{A}\mathbf{u}+\mathbf{t}$, where $\mathbf{u}$ is a column vector $[u_1;u_2]$ and the matrix $\mathcal{A}$ is composed of two lattice vectors $[\mathbf{A}_1,\mathbf{A}_2]$. The point $\mathcal{R}(\varphi)\mathcal{A}\mathbf{u}+\mathbf{t}$ projects to a point at the lower layer $s_1\mathbf{A}_1+s_2\mathbf{A}_2=\mathcal{A}\mathbf{s}$. Then we have 

\begin{equation}
\begin{aligned}
\mathbf{u}=\mathcal{A}^{-1}\mathcal{R}(-\varphi)\mathcal{A}\mathbf{s}-\mathcal{A}^{-1}\mathcal{R}(-\varphi)\mathbf{t}.
\end{aligned}
\end{equation}

The rotation leads to the appearing of Mori\'{e} pattern. The special points of the Mori\'{e} pattern, for example the skyrmion centers, have special local stacking structures. It can be described by special values of $\mathbf{c}=\mathbf{u}-\mathbf{s}$. Therefore, the following equation describes the motion of a skyrmion.

\begin{equation}
\begin{aligned}
\mathbf{c}=\mathcal{A}^{-1}\mathcal{R}(-\varphi)\mathcal{A}\mathbf{s}-\mathcal{A}^{-1}\mathcal{R}(-\varphi)\mathbf{t}-\mathbf{s}.
\end{aligned}
\end{equation}

Lets first consider the translation. In this case,  $\mathbf{s}$ and $\mathbf{t}$ are variables, while $\mathbf{c}$ and the twist angle keep constants, 

\begin{equation}
\begin{aligned}
\mathcal{A}^{-1}\mathcal{R}(-\varphi)\mathcal{A}d\mathbf{s}-\mathcal{A}^{-1}\mathcal{R}(-\varphi)d\mathbf{t}-d\mathbf{s}=0.
\end{aligned}
\end{equation}

Then, we have

\begin{equation}
\begin{aligned}
d\mathbf{t}=(1-\mathcal{R}(\varphi))\mathcal{A}d\mathbf{s}.
\end{aligned}
\end{equation}

In the condition $\varphi\ll\pi$, we have 

\begin{equation}
\begin{aligned}
1-\mathcal{R}(\varphi)\approx\mathcal{K}(\varphi):=\begin{bmatrix}
0        &  \varphi  \\
-\varphi  &  0
\end{bmatrix}.
\end{aligned}
\end{equation}

The matrix is anti-symmetric $\mathcal{K}^T(\varphi)=-\mathcal{K}(\varphi)$, and $\mathcal{K}^T(\varphi)\mathcal{K}(\varphi)=\varphi^2\mathcal{I}$. The $\mathcal{I}$ here is the unitary matrix. The motion of the Mori\'{e} pattern ($\mathcal{A}d\mathbf{s}$) and the translation of the upper layer ($d\mathbf{t}$) are perpendicular with each other, since their inner product is zero.

\begin{equation}
\begin{aligned}
(\mathcal{A}d\mathbf{s})^Td\mathbf{t}&=d\mathbf{s}^T\mathcal{A}^T\mathcal{K}(\varphi)\mathcal{A}d\mathbf{s} \\
&=-d\mathbf{t}^T\mathcal{A}d\mathbf{s}\\
&=0.
\end{aligned}
\end{equation}

And their ratio of speed is

\begin{equation}
\begin{aligned}
\frac{\vert\mathcal{A}d\mathbf{s}\vert }{\vert d\mathbf{t}\vert }\approx\sqrt{ \frac{d\mathbf{s}^T\mathcal{A}^T\mathcal{A}d\mathbf{s}}{d\mathbf{s}^T\mathcal{A}^T\mathcal{K}^T(\varphi)\mathcal{K}(\varphi)\mathcal{A}d\mathbf{s}}  }=\frac{1}{\varphi}.
\end{aligned}
\end{equation}

Thus, a small amount of layer translation will lead the skyrmion to move a long distance. It can be used to tune the position of skyrmions.

Lets further consider the change of twist angle. We will discuss the non-translation ($\mathbf{t}=0$) case, where the $\varphi$ and $\mathbf{s}$ are changable variables.
From Eq.17, we know that

\begin{equation}
\begin{aligned}
(1-\mathcal{R}(\varphi))\mathcal{A}d\mathbf{s}&=\frac{\partial\mathcal{R}(\varphi)}{\partial\varphi}\mathcal{A}\mathbf{s}d\varphi\\
&=\mathcal{R}(\frac{\pi}{2}+\varphi)\mathcal{A}\mathbf{s}d\varphi.
\end{aligned}
\end{equation}

In the small twist angle condition, we have  

\begin{equation}
\begin{aligned}
\mathcal{A}d\mathbf{s}&\approx(1-\mathcal{R}(-\varphi))\mathcal{R}(\frac{\pi}{2}+\varphi)\mathcal{A}\mathbf{s}d\varphi\\
&=\mathcal{R}(\frac{\pi}{2})(\mathcal{R}(\varphi)-1)\mathcal{A}\mathbf{s}d\varphi\\
&\approx -\varphi\mathcal{A}\mathbf{s}d\varphi\\
&=-\frac{\mathcal{A}\mathbf{s}}{2}d\varphi^2.
\end{aligned}
\end{equation}

From this equation, we know that the translation of skyrmion center $\mathcal{A}d\mathbf{s}$ is propotional to its position vector $\mathcal{A}\mathbf{s}$. The skyrmions get close to each other when $\vert \varphi\vert $ increases, and get far from each other when  $\vert \varphi\vert $ decreases, and the period of Mori\'{e} pattern and the skyrmion size also decrease or increase accordingly. 

These discussions are based on the condition $\varphi\ll\pi$. It seems conflicting with $\varphi\sim60^o$ in the T-TBCI. Actually, we can redefine the lattice vectors of a $\pi/3$-rotated CrI$_3$ layer to be same with that of the unrotated CrI$_3$ layer, and redefine the twist angle as $\varphi'=\varphi-60^o$. Then the above discussions are still valid for T-TBCI.

\bibliography{zSI}